# Ant Based Adaptive Multicast Routing Protocol (AAMRP) for Mobile Ad Hoc Networks

A. Sabari
Asst.Professor, Department of Information Technology
K.S.Rangasamy College of Technology
Tiruchengode, India.
asabariphd@yahoo.com

K.Duraiswamy
Professor, Dept of Computer Science and Engineering
K.S.Rangasamy College of Technology
Tiruchengode, India.

*Abstract*—**Multicasting is effective when its group members are sparse and the speed is low. On the other hand, broadcasting is effective when the group members dense and the speed are high. Since mobile ad hoc networks are highly dynamic in nature, either of the above two strategies can be adopted at different scenarios. In this paper, we propose an ant agent based adaptive, multicast protocol that exploits group members' desire to simplify multicast routing and invoke broadcast operations in appropriate localized regimes. By reducing the number of group members that participate in the construction of the multicast structure and by providing robustness to mobility by performing broadcasts in densely clustered local regions, the proposed protocol achieves packet delivery statistics that are comparable to that with a pure multicast protocol but with significantly lower overheads. By our simulation results, we show that our proposed protocol achieves increased Packet Delivery Fraction (PDF) with reduced overhead and routing load.**

1. INTRODUCTION

Multicasting is the transmission of data grams to a group of hosts identified by a single destination address. Multicasting is intended for group-oriented computing. There are more and more applications where one-to-many dissemination is necessary. The multicast service is critical in applications characterized by the close collaboration of teams (e.g. rescue patrol, battalion, scientists, etc) with requirements for audio and video conferencing and sharing of text and images. The use of multicasting within a network has many benefits. Multicasting reduces the communication costs for applications that send the same data to multiple recipients. Instead of sending via multiple unicasts, multicasting minimizes the link bandwidth consumption, sender and router processing, and delivery delay.

Maintaining group membership information and building optimal multicast trees is challenging even in wired networks However, nodes are increasingly mobile. One particularly challenging environment for multicast is a mobile ad-hoc network MANET.

A MANET consists of a dynamic collection of nodes with sometimes rapidly changing multi-hop topologies that is composed of relatively low-bandwidth wireless links. Since each node has a limited transmission range, not all messages may reach all the intended hosts. To provide communication through the whole network, a source-to-destination path could pass through several intermediate neighbor nodes. Unlike typical wire line routing protocols, ad hoc routing protocols must address a diverse range of issues. The network topology can change randomly and rapidly, at unpredictable times.

Since wireless links generally have lower capacity, congestion is typically the norm rather than the exception.

It is observed that typical multicast protocols are geared toward and optimized for particular scenarios. Therefore, when they are deployed in different scenarios, their performance may vary significantly. Furthermore, they may incur unreasonable amounts of overheads in certain scenarios. The creation and maintenance of the multicast structure could be heavyweight as their operations require control messages to be exchanged among the constituent nodes in the network. In cases of high mobility, wherein the constructed multicast structure tends to stale fairly quickly, there is a need for the periodic invocation of control messages with high frequency.

Broadcasting provides several basic advantages. First, it does not require the creation of any delivery structure. Second, there is a natural redundancy in broadcasting due to multiple rebroadcast nodes. This redundancy provides extra robustness in conditions of mobility. Therefore, broadcasting is preferable for use in the scenarios with large group members or in case of high mobility. On the negative side, broadcasting would attempt to deliver the packet to all the nodes in the network regardless of who the intended recipients are. This property of broadcasting leads to many redundant data transmissions and renders it an unsuitable choice in scenarios with a small number of group members.

Ant agent based adaptive, multicast protocol that exploits group members' desire to simplify multicast routing and invoke broadcast operations in appropriate localized regimes has been proposed. By reducing the number of group members that participate in the construction of the multicast structure and by providing robustness to mobility by performing broadcasts in densely clustered local regions, the proposed protocol achieves packet delivery statistics that are comparable to that with a pure multicast protocol but with significantly lower overheads.

II. EXISTING WORK ON MANET MULTICAST ROUTING

*A. Ant Based MANET Multicast Routing Algorithms*

Ant colony optimization is a probabilistic technique for solving computational problems which can be reduced to find the good paths through graphs. Ants are used as the agents and the routing is on basis of the food searching behavior of the real ants. These agents are divided into forward and backward ants. The sender to the neighbor nodes broadcasts the forward ants. The backward ants utilize the useful information like end-to-end delay, number of hops gathered by the forward ants on their trip from source to the destination.

It has been experimentally observed that ants in a colony can converge on moving over the shortest among different paths connecting their nest to a source of food. The main catalyst of this colony-level shortest path behavior is the use of a volatile chemical substance called pheromone: ants moving between the nest and a food source deposit pheromone, and preferentially move in the direction of areas of higher pheromone intensity. Shorter paths can be completed quicker and more frequently by the ants, and will therefore be marked with higher pheromone intensity. These paths will therefore attract more ants, which will in turn increase the pheromone level, until there is convergence of the majority of the ants onto the shortest path. The local intensity of the pheromone field, which is the overall result of the repeated and concurrent path sampling experiences of the ants, encodes a spatially distributed measure of goodness associated with each possible move.

Lin Huang, Haishan Han and Jian Hou [1], have proposed an efficient algorithm for generating a low-cost multicast routing, subject to delay constraints (ASDLMA), based on Ant System algorithm. When an ant on the Graph moves from a node to other node depend on the corresponding probabilities function, and update the pheromone on Graph when every iteration finished. Simulation results show our algorithm has features of well performance of cost, fast convergence and stable delay.

Diego Pinto, Benjamí n Barán and Ramón Fabregat [2], have presented a new multi objective algorithm based on ant colonies, which is used in the construction of the multicast tree for data transmission in a computer network. The proposed algorithm simultaneously optimizes cost of the multicast tree, average delay and maximum end-to-end delay.

Hua Wang, Zhao Shi, Shuai Li [3], have proposed an ant colony algorithm with orientation factor and applies it to multicast routing problem with the constraints of delay variation bound. The orientation factor enables the ant to get rid of the initial blindness when searching paths, makes use of the search results and reduces the misguiding effect of pheromone on irrelevant paths, thus overcoming the drawbacks of slow convergence existing in the basic ant colony algorithm, increasing the speed of convergence and speeding up the finding of feasible solution to the problem.

Diego Pinto, and Benjamín Barán [4], have proposed a new approach for the resolutions of Multi-Objective Problems (MOPs) inspired in Max-Min Ant System (MMAS), where Ant Colony Optimization (ACO) already resolve the single-objective combinatorial problems. A multicast traffic-engineering problem was solved using the proposed approach as well as a Multiobjective Multicast Algorithm (MMA), a Multi-objective evolutionary algorithm (MOEA) specially designed for that multicast problem.

Zeyad M. Alfawaer, GuiWei Hua, and Noraziah Ahmed [5], have introduced MANHSI (Multicast for Ad hoc Network with hybrid Swarm Intelligence) protocol, which relies on a swarm intelligence based optimization technique to learn and discover efficient multicast connectivity. The proposed protocol instances that it can quickly and efficiently establish initial multicast connectivity and/or improved the resulting connectivity via different optimization techniques

*B, Other MANET Multicast Routing Algorithms*

Existing MANET multicast routing algorithms use a hybrid of link state and distance vector algorithms to create a source tree. As mentioned before, multicast is a relatively new topic in the field of MANETs, and only a few algorithms have been proposed. Examples of some MANET multicast algorithms are

**Position-Based Multicast (PBM)** [6]
PBM a forwarding node uses information about the positions of the destinations and its own neighbors to determine the next hops that the packet should be forwarded to. It is thus very well suited for highly dynamic networks.

**Application Layer Multicast with Network Layer Support (APPMULTICAST)** [7]
APPMULTICAST is an efficient application layer multicast solution suitable for medium mobility applications in MANET environment. We use network layer support to build the overlay topology closer to the actual network topology.

**Robust Demand-driven Video Multicast Routing (RDVMR)** [8]
Our protocol uses a novel path based Steiner tree heuristic to reduce the number of forwarders in each tree, and constructs multiple trees in parallel with reduced number of common nodes among them.

**Application Layer Multicast Algorithm** [9]
ALMA constructs an overlay multicast tree of logical links between the group members. It is receiver-driven, flexible and highly adaptive

**Multicast Tree Algorithm** [10]
The algorithm utilizes shortest path information, which can be obtained from unicast routing tables for fully distributed implementation.

**Progressively Adapted Sub-Tree in Dynamic Mesh(PAST-DM)** [11]
In PAST-DM the virtual mesh topology gradually adapts to the changes of underlying network topology in a fully distributed manner with minimum control cost. The multicast tree for packet delivery is also progressively adjusted according to the current topology.

**Multicast AODV (MAODV)** [12],
In MAODV, control of the multicast tree is distributed so that there is no single point of failure.

**On-Demand Multicast Routing Protocol (ODMRP)** [13],
ODMRP applies on-demand routing techniques to avoid channel overhead and improve scalability. It uses the concept of forwarding group, a set of nodes responsible for forwarding multicast data on shortest paths between any member pairs, to build a forwarding mesh for each multicast group.

**Multicast Zone Routing (MZR)** [14],
MZR is a source-initiated on demand protocol, in which a multicast delivery tree is created using a concept called the zone routing mechanism. It is a source tree based protocol and does not depend on any underlying unicast protocol.

**Multicast OLSR (MOLSR)** [15],

MOLSR is in charge of building a multicast structure in order to route multicast traffic in an ad-hoc network. MOLSR is designed for mobile multicast routers, and works in a heterogeneous network composed of simple unicast OLSR routers, MOLSR routers and hosts.

**Ad hoc Multicast Routing (AM Route)** [16],

AM Route presents a novel approach for robust IP Multicast in mobile adhoc networks by exploiting user multicast trees and dynamic logical cores. It creates a bi directional, shared tree for data distribution using only group senders and receivers as tree nodes.

**Core-Assisted Mesh Protocol (CAMP)** [17],

CAMP) is introduced for multicast routing in ad-hoc networks. CAMP generalizes the notion of core-based trees introduced for internet multicasting into multicast meshes that have much richer connectivity than trees.

**Adaptive Demand Driven Multicast Routing (ADMR)** [18],

ADMR attempts to reduce as much as possible any non on demand components within the protocol. Multicast routing state is dynamically established and maintained only for active groups and only in nodes located between multicast senders and receivers.

**Lightweight Adaptive Multicast Algorithm (LAM)** [19],

LAM offers the basic functionalities of a multicasting protocol that is to construct and maintain routing path tree(s) inter-connecting members of the multicasting group so that all multicast data can be forwarded using only the tree paths.

**Multicast Core Extraction Distributed Ad-Hoc Routing (MCEDAR)** [20].

MCEDAR is an extension to the CEDAR architecture and provides the robustness of mesh based routing protocols and the approximates the efficiency of tree based forwarding protocols

**Protocol for unified multicasting through announcements (PUMA)** [21]

PUMA in ad-hoc networks establishes and maintains a shared mesh for each multicast group, without requiring a unicast routing protocol or the pre assignment of cores to groups.

### III. ANT BASED ADAPTIVE MULTICAST ROUTING PROTOCOL (AAMRP)

First, a simple broadcast scheme can significantly reduce the control overhead in scenarios wherein the density of group members is high. Second, many current protocols cannot adapt to local variations in network properties. Most of these protocols have static, globally predefined parameters that cannot be adjusted dynamically within localized regimes. Our objective then is to design a new protocol that

- Exploits the advantages of broadcasting in high densities and
- Provides localized flexibility in response to changing network conditions.

AAMRP dynamically identifies and organizes the group members into clusters which correspond to areas of high group member affinity. In each of these "dense" neighborhoods, one of the group members is selected to be a cluster leader. Cluster leaders have two main functions:

- They establish a sparse multicast structure among themselves and the source, and
- They use broadcasting (with adaptive scope) to deliver the packets to other group members in their cluster.

#### A. Algorithm Description

It constructs a 2-tier hierarchical structure, where the upper tier is formed by a multicast source and cluster leaders that represent groups of multicast members that form a cluster, and the lower tier consists of the members in a cluster. Since each cluster demonstrates a high density of group members, a cluster leader simply invokes an adaptive localized broadcast within its cluster to disseminate multicast packets received from the source. This would reduce the consumed overhead while ensuring efficient data delivery.

#### B. Construction of the Multicast Structure

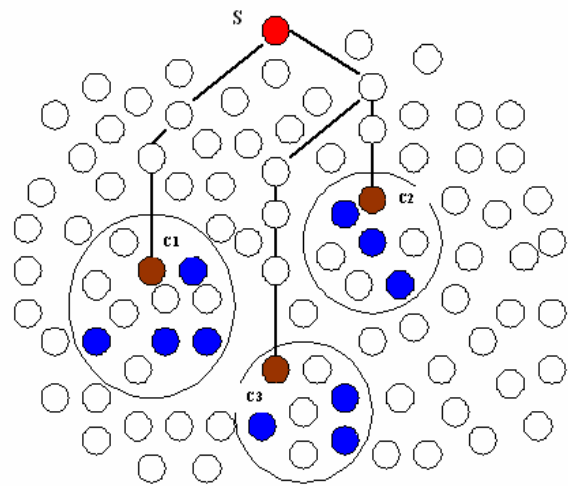

**Figure1.** Proposed Multicast Architecture

#### 3.2.1 Determination of Group Members

Each group member in AAMRP can be in 3 states. It can be in a temporary mode wherein it is JOINING the session, it can be a cluster LEADER, or it can simply be the MEMBER of a cluster leader.

Each node maintains a Group Member Table (GMTable) which contains the information of the joining group members. The information maintained in this table is obtained by means of the ADVERTISE and the LEADER messages. Each cluster leader maintains a Cluster Member Table (CMTable), which contains information of all the cluster group members that are associated with the cluster leader. The information maintained in this table is obtained via the reception of MEMBER messages that are sent out by each cluster member.

**1. Discovery Phase.** In this phase, the joining node discovers the other joining group members and cluster leaders

in its vicinity. When a node decides to join a multicast group, it enters this phase and informs its presence to its k-hop neighborhood by broadcasting a JOIN message. The JOIN message contains the address, multicast address, hopcount information. On receiving a unique JOIN message, each node updates its GMTable as per the contents of the message. After this phase, each joining node would have obtained the k-hop local topology information in their GMTables, which may be used to determine the cluster leaders in the decision phase. When the connection to the cluster leader is lost, this phase is executed again.

**2. Leader Election Phase**. In this phase, the joining node elects itself as the cluster leader for its k-hop neighborhood, if it cannot still find any cluster leader in its vicinity, after the discovery phase. If the inter-connectivity of a node is highest when compared to its k-hop neighbors, it will elect itself as a cluster leader and serve a cluster. It then changes its role to LEADER and broadcasts a LEADER message containing its address, multicast-address, connectivity and hop count information. Nodes that are within the broadcast range of the LEADER message, update their GMTable to reflect the contents of the message.

A cluster leader is considered to be best, when it has the shortest distance, highest connectivity and highest node Id. Among several LEADER messages received, the joining node will select the best cluster leader by sending a MEMBER message containing its address, multicast-address and hop count information to the selected cluster leader. This is to inform the cluster leader that it is going to join the cluster. The cluster leader would then update its CMTable accordingly.

After the completion of the above phases, a joining node must either become a cluster leader or a child of a cluster leader. From then on, each cluster formed becomes a single routing entity as represented by its cluster leader. Only the relatively small number of cluster leaders will then participate in the construction and maintenance of the multicast structure.

**3.2.2 Creation of Upper Tier Multicast Structure**
The multicast source broadcasts a MCAST-REQ message containing its address and multicast-group to the network, periodically. Intermediate nodes forward the unique MCAST-REQ messages further and the backward ant agent stores the path toward the source. When a cluster leader receives the MCAST-REQ message, it sends a MCAST-REP message back to the source, through the path established by the backward ant agents. The forwarding nodes along the path toward the source, are represented by the (source, multicast-group) attribute pair. Then data packets are multicast from the source to the cluster leaders through the tree constructed by the ant agents. When the forwarding nodes receive MCAST-REP from more than one cluster leader, they forward the packets to their multiple children on the tree.

**Step1: Backup-paths-set**
For each destination node $m_i \in M$, Dijkstra $K$ shortest path algorithm is used to compute the least-cost paths from s to m to construct a backup-path set. Let $P_i$ be paths set for destination node $i$:

$$P_i = \{P_i^1, \ldots P_i^j, \ldots P_i^k\} \quad (1)$$

Where $P_i^j$ is the jth path for destination node $i$

If the delay constraint is violated by some of the trees, then the cost is to be increased, so that it is likely to be rejected.

**Step2: Tree Formation**
In this algorithm, $a$ multicast tree $T$ is represented as an array of m elements,

$$T = \{P_1, P_2, \ldots P_m\} \quad (2)$$

Where $P_i = P(s, m_i)$, is the path set selected from (1), $s$ is the source and $m_i$ is the destination.

**Step3: Path selection**
When an ant moves from the node $i$ to the next node $j$, the probability function of the ant choosing node $j$ as the next node as follows:

$$f_{ij} = \begin{cases} \dfrac{[T_{ij}]^\alpha [\eta_{ij}]^P}{\sum u \in N_h(t)^{[T_{iu}]\alpha[\eta_{iu}]^\beta}} & \text{if } j \in N_h(i) \\ 0 & \text{otherwise} \end{cases} \quad (3)$$

$\alpha$ and $\beta$ are the relative importance of pheromone strength and the distance between nodes that affect an ant's judgment when choosing the next node to select.

**Step4: Pheromone update**
The pheromone trail associated to every edge is evaporated by reducing all pheromones by a constant factor:

$$\tau_{ij} \leftarrow (1-P)\ \tau_{ij} \quad (4)$$

Where $p \in (0,1)$ is the evaporation rate. Next, each ant retracts the path it has followed and deposits an amount of pheromone $\Delta \tau_{ij}^h$ on each traversed connection

$$\tau_{ij} \leftarrow \tau_{ij} + \Delta \tau_{ij}^h, \quad \alpha_{ij} \in \text{Sh} \quad (5)$$

The pheromone on a connective path $(i.j)$ left by the mth ant is the inverse of the total length traveled by the ant in a particular cycle. The formula is as follows

$$\tau_{ij}^h = Q/Lm$$

In the above formula,

$Q$ is a constant, and $Lm = (C_j - Ci)$, where $c_i$ is cost of sub multicast tree node $i$ and $c_j$ is cost of sub multicast tree node j. To avoid the situation of $c_i = c_j$ compute

$$Lm = (C_j - C_i)^2 + 1 \qquad (6)$$

**Step5: Stopping criterion**
The stopping criterion of the algorithm could be specified by a maximum number of iterations or a specified CPU time limit

**3.2.3 Adaptive Broadcast within a Cluster**
When the cluster leader receives a data packet from the multicast source, it broadcast it to the group members within its cluster. Here adaptive broadcast is performed, since the maximum broadcast range is depending on the furthest child of the cluster leader. (i.e.), the broadcast range can be reduced as per the distance of the furthest child, stored in the CMTable. Due this adaptive broadcast, redundant and unwanted transmissions of data can be reduced.

*C. Joining a Multicast Group*

To join a multicast group, the state of the node should be either a cluster leader or cluster member. The process of joining a multicast group is described below.

When a node decides to join a multicast group, it simply changes its role to JOINING and enters the discovery and leader election phase as described in the previous section. If the joining node has cluster leaders in its k-hop vicinity, it would possibly receive LEADER messages before entering the leader election phase. In this case, the joining node will simply pick the best cluster leader to join as described in previous section. If the joining node has no cluster leader present in its vicinity and its connectivity is the highest as compared to its k-hop neighbors, it will become a cluster leader and serve a cluster.

*D. Leaving a Multicast Group*

Group members could leave a multicast group at any time. A group member that has the state of MEMBER simply stops sending the MEMBER message to its cluster leader.

When a cluster leader decides to leave the multicast group, it simply stops transmitting the LEADER message. Cluster members, upon discovering the absence of a leader, will first try to quickly rejoin another cluster by looking for other leaders in their GMTable. If no cluster leader is present in a member's vicinity, the cluster member will switch its role to JOINING and invoke the discovery and decision phases to find another cluster or to become a cluster leader as described in Section 3.2.1.

*E. Maintaining the Multicast Structure*

**3.5.1 Functions of Source**
The upper tier multicast structure is refreshed by the source periodically, by exchanging the MCAST-REQ and MCAST-REP messages as described in section 3.2.2. When the paths are changed due to mobility, the old routes are deleted and new routes are added.

**3.5.2 Functions of Cluster Leader and Member**
Each cluster leader periodically broadcasts a LEADER message to its cluster, to notify its presence to the cluster members. Also this message can be used to invite new group members that are not currently associated with the cluster. Each cluster member estimates the distance from the leader to itself, and send this distance information to the leader through a unicast MEMBER message, periodically. The frequency of these unicast updates from a member depends on this distance of the member from the leader. (i.e., if the cluster member is closer to its leader, it is less possible that it is out of range).Using this information, the cluster leader dynamically adjust the range of the local broadcast such that,

Range R = { 2 hops, if (Nocm + Nncm) > T
          { 1 hop, otherwise

where Nocm is the number of old clusters members of the cluster , Nncm is the number of new cluster members of the cluster and T is the system defined threshold value.

When the broadcast range R is set to 1 hop, unicast transmission will take place from the source to the associated cluster members through the leader.

A cluster member may switch to a new cluster , when it detects that the new cluster leader is closer to its current leader. It is done by overhearing the LEADER message of other cluster leaders.

IV. EXPERIMENTAL RESULTS

*A. Simulation Setup*

NS2 is used to simulate the proposed algorithm. In our simulation, the channel capacity of mobile hosts is set to the same value: 2 Mbps. The distributed coordination function (DCF) of IEEE 802.11 for wireless LANs as the MAC layer protocol is used. It has the functionality to notify the network layer about link breakage.

In the simulation, mobile nodes move in a 600 meter x 600 meter rectangular region for 50 seconds simulation time. Initial locations and movements of the nodes are obtained using the random waypoint (RWP) model of NS2. I assume each node moves independently with the same average speed. All nodes have the same transmission range of 250 meters. In this mobility model, a node randomly selects a destination from the physical terrain. It moves in the direction of the destination in a speed uniformly chosen between the minimal speed and maximal speed. After it reaches its destination, the node stays there for a pause time and then moves again.

In the simulation, the maximal speed is 10 m/s. and pause time is 5sec. We vary the number of nodes as 25, 50, 75 and 100, to investigate the performance influence of different topologies. The simulated traffic is Constant Bit Rate (CBR). For each scenario, ten runs with different random seeds were conducted and the results were averaged.

AAMRP is compared with our previous protocol Ant Based Multicast Routing (AMR) [22] and Multicast for Ad hoc Network with Hybrid Swarm Intelligence (MANHSI) [5]. The evaluation is mainly based on performance according to the following metrics:

**Control overhead**: The control overhead is defined as the total number of routing control packets received.

**Routing Load:** The normalized routing load is the ratio of no. of routing packets and the total no. of packets received.
**End to End Delay:** It is average end-to-end-delay of the transmission.
**Packet Delivery Fraction**: It is the ratio of the fraction of packets received successfully and the total no. of packets sent
The simulation results are presented in table1 and table2 of next section.

*B. Results*

**A. Effect of Network Size**
In this experiment, we vary the network size by varying the number of nodes as 25,50,75 and 100.

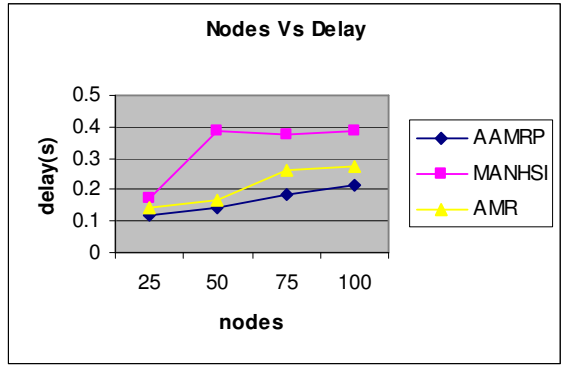

Figure 2: Nodes Vs End-to-End Delay

Figure 2 shows that the end-to-end delay of the proposed AAMRP is significantly less when compared with AMR and MANHSI, since the multicast tree formation involves less overhead when compared with other algorithms. In the figure we can see that, as the network size increases, the corresponding delay also increases.

Figure 3 shows the packet delivery fraction (PDF) of all the protocols. From the figure, we can see that the PDF of AAMRP is slightly more than AMR and significantly more than MANHSI, since the mobility induced errors are minimized in AAMRP.

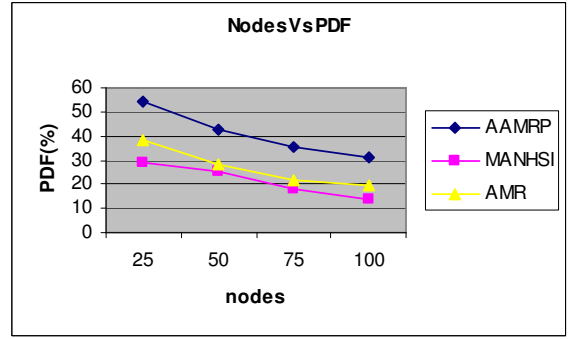

Figure 3: Nodes Vs Packet Delivery Fraction

Figure 4 shows the control overhead occurred in all the protocols. From the figure, we can observe that the control overhead increases when network size grows. Since less control message is exchanged in AAMRP, its overhead is less than that of AMR and MANHSI.

TABLE: 1 EFFECT OF VARYING NODE

| Nodes | Overhead | | | Load | | | Delay | | | PDFraction | | |
|---|---|---|---|---|---|---|---|---|---|---|---|---|
| | AAMRP | AMR | MANSHI | AAMRP | AMR | MANSHI | AAMRP | AMR | MANSHI | AAMRP | AMR | MANSHI |
| 25 | 573 | 745 | 823 | 0.1100 | 0.23714 | 0.46079 | 0.121 | 0.1458 | 0.1738 | 54.2143 | 38.5447 | 28.9941 |
| 50 | 750 | 850 | 963 | 0.3602 | 0.5501 | 0.82476 | 0.1409 | 0.1674 | 0.3852 | 42.4902 | 28.474 | 25.5090 |
| 75 | 766 | 861 | 1109 | 0.6712 | 0.7284 | 1.49371 | 0.1827 | 0.2620 | 0.3748 | 35.5232 | 21.784 | 17.8556 |
| 100 | 1122 | 1274 | 1865 | 0.6935 | 0.8581 | 1.87528 | 0.2126 | 0.2723 | 0.3864 | 31.4371 | 19.231 | 13.5428 |

TABLE: 2 EFFECT OF VARYING SIZE

| Group Size | Overhead | | | Load | | | Delay | | | PDFraction | | |
|---|---|---|---|---|---|---|---|---|---|---|---|---|
| | AAMRP | AMR | MANSHI | AAMRP | AMR | MANSHI | AAMRP | AMR | MANSHI | AAMRP | AMR | MANSHI |
| 1 | 1567 | 1714 | 1812 | 0.3571 | 0.3942 | 0.4274 | 0.0178 | 0.01937 | 0.02049 | 87.485 | 87.5848 | 82.3141 |
| 2 | 2466 | 3128 | 4194 | 0.6712 | 1.4342 | 1.7478 | 0.04327 | 0.06562 | 0.07684 | 52.523 | 52.5232 | 50.1742 |
| 3 | 4685 | 5234 | 5814 | 1.3635 | 1.7218 | 2.1890 | 0.06133 | 0.07145 | 0.09182 | 38.263 | 38.2628 | 35.1743 |
| 4 | 5473 | 6452 | 7163 | 1.6814 | 2.1098 | 2.9440 | 0.08419 | 0.09215 | 0.10942 | 29.685 | 29.6854 | 27.5673 |

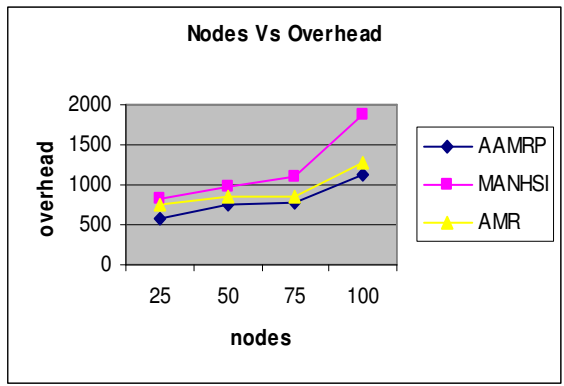

Figure 4: Nodes Vs Overhead

Figure 4 shows the normalized routing load of both the protocols. From the figure, we can see that the routing load is significantly less for AAMRP, when compared to MANHSI.

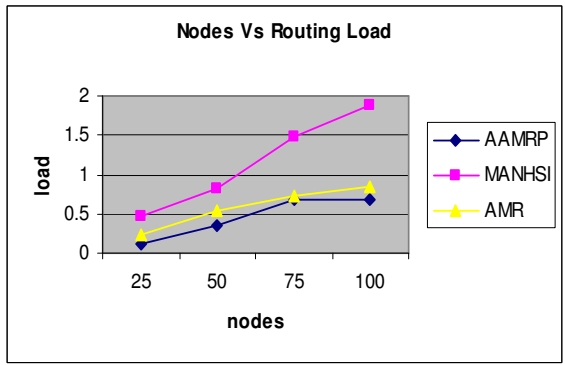

Figure 5: Nodes Vs Routing Load

**B. Effect Of Group Size**
In this experiment, we vary the group size as 1,2,3 and 4.

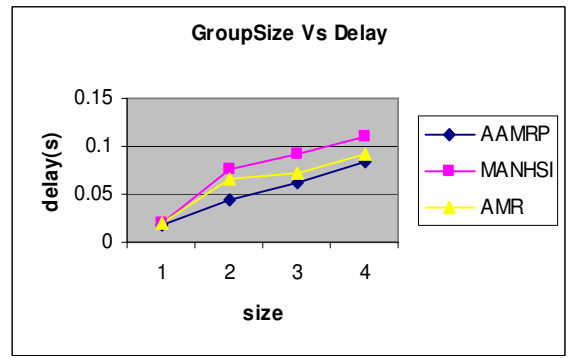

Figure 6: Group Size Vs End-to-End Delay

Figure 6 shows that the end-to-end delay of the proposed AAMRP is significantly less when compared with AMR and MANHSI, since the multicast tree formation involves less overhead when compared with MANHSI algorithm. In the figure we can see that, as the group size increases, the corresponding delay also increases.

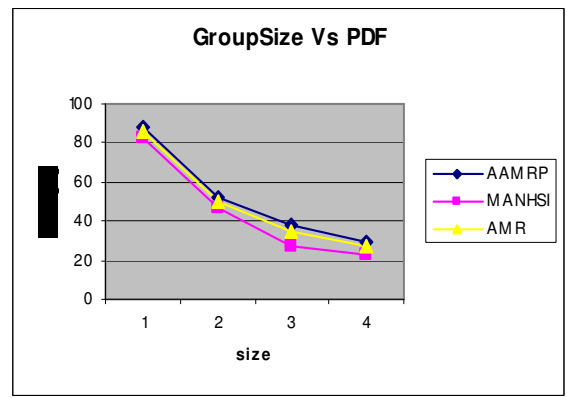

Figure 7: Group Size Vs Packet Delivery Fraction

Figure 7 shows the packet delivery fraction (PDF) of all the protocols. From the figure, we can see that the PDF of AAMRP is slightly more than AMR and significantly more than MANHSI, since the mobility induced errors are minimized in AAMRP.

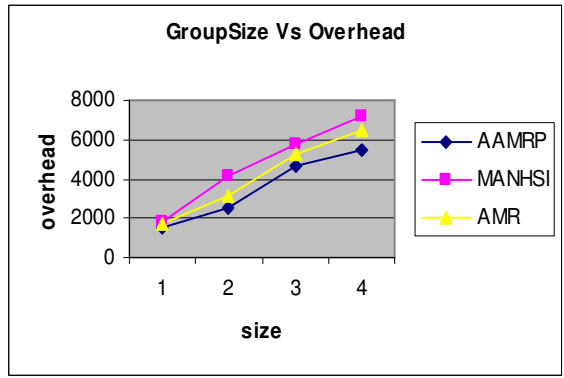

Figure 8:Group Size Vs Overhead

Figure 8 shows the control overhead occurred in all the protocols. From the figure, we can observe that the control overhead increases when group size grows. Since less control message is exchanged in AAMRP, its overhead is less than that of AMR and MANHSI.

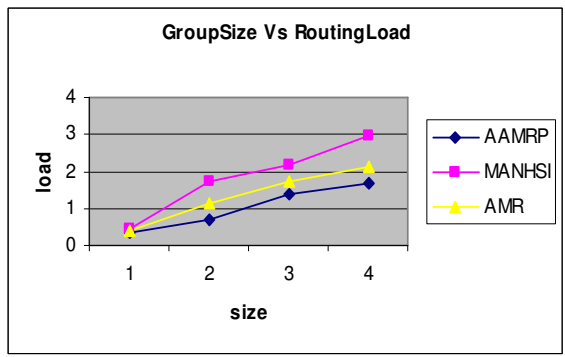

Figure 9:Nodes Vs Routing Load

Figure 9 shows the normalized routing load of both all protocols. From the figure, we can see that the routing load is significantly less for AAMRP, when compared to AMR and MANHSI.

## V. CONCLUSION

In this paper, we have proposed ant agent based adaptive multicast protocol which combines the positive aspects both multicasting and broadcasting .It exploits group members' desire to simplify multicast routing and invoke broadcast operations in appropriate localized regimes. By reducing the number of group members that participate in the construction of the multicast structure and by providing robustness to mobility by performing broadcasts in densely clustered local regions, the proposed protocol achieves packet delivery statistics that are comparable to that with a pure multicast protocol but with significantly lower overheads. By our simulation results, we have shown that our proposed protocol achieves increased PDF with reduced overhead and routing load.

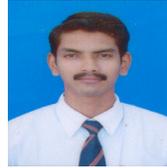

Sabari.A received the B.E. degree in Computer Science and Engineering from the University of Madras in 2000, and the M.Tech. degree in Computer Science and Engineering from the Visveswaraiah Technological University, Belgaum in 2002. His research activity includes multicasting and mobile ad hoc networks. He is currently working in K.S.Rangasamy College of Technology, Tiruchengode as Assistant Professor of Information Technology Department since 2002. He is a life member of ISTE and CSI.

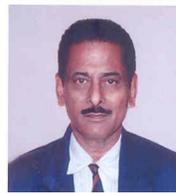

Dr. K.Duraiswamy (SM) received his B.E. degree in Electrical and Electronics Engineering from P.S.G. College of Technology, Coimbatore, Tamil Nadu in 1965 and M.Sc.(Engg) degree from P.S.G. College of Technology, Coimbatore, Tamil Nadu in 1968 and Ph.D. from Anna University, Chennai in 1986. From 1965 to 1966 he was in Electricity Board. From 1968 to 1970 he was working in ACCET, Karaikudi, India. From 1970 to 1983, he was working in Government College of Engineering, Salem. From 1983 to 1995, he was with Government College of Technology, Coimbatore as Professor. From 1995 to 2005 he was working as Principal at K.S. Rangasamy College of Technology, Tiruchengode and presently he is serving as Dean in the same institution. He is interested in Digital Image Processing, Computer Architecture and Compiler Design. He received 7 years Long Service Gold Medal for NCC. He is a life member in ISTE, Senior member in IEEE and a member of CSI.